\documentclass[11pt]{article}

\usepackage{graphicx}
\usepackage[T1]{fontenc}
\usepackage{fleqn}
\usepackage{amsmath}

\begin{document}
\begin{normalsize}

\begin{center}

{\bf Eta Photoproduction on the Neutron at GRAAL:\\
       Measurement of the Differential Cross Section}
       
\end{center}

\vspace{0.1in}

\begin{center}

\noindent \small {D.~Rebreyend~$^1$,
O.~Bartalini~$^2$,
V.~Bellini~$^3$,
J.P.~Bocquet~$^1$,
M.~Castoldi~$^4$,
A.~D'Angelo~$^2$,
J.-P.~Didelez~$^5$,
R.~Di~Salvo~$^2$,
A.~Fantini~$^2$,
G.~Gervino~$^6$,
F.~Ghio~$^7$,
B.~Girolami~$^7$,
A.~Giusa~$^3$,
M.~Guidal~$^5$,
E.~Hourany~$^5$,
V.~Kouznetsov~$^8$,
R.~Kunne~$^5$,
A.~Lapik~$^8$,
P.~Levi~Sandri~$^9$,
A.~Lleres~$^1$,
D.~Moricciani~$^2$,
V.~Nedorezov~$^8$,
G.~Russo~$^3$,
N.~Rudnev~$^8$,
C.~Schaerf~$^2$,
M.-L.~Sperduto~$^3$,
M.-C.~Sutera~$^3$,
A.~Turinge~$^{10}$\\
(The GRAAL collaboration)}

\end{center}

\vspace{0.1in} 

\begin{center}

\noindent \small \it {$^1$ IN2P3, Laboratory for Subatomic Physics and Cosmology, 38026 Grenoble, France\\
$^2$ INFN sezione di Roma II and Universit\`a di Roma ``Tor Vergata", 00133 Roma, Italy\\
$^3$ INFN sezione di Catania and Universit\`a di Catania, 95100 Catania, Italy\\
$^4$ INFN sezione di Genova and Universit\`a di Genova, 16146 Genova, Italy\\
$^5$ IN2P3, Institut de Physique Nucl\'eaire, 91406 Orsay, France\\
$^6$ INFN sezione di Torino  and Universit\`a di Torino, 10125 Torino, Italy\\
$^7$ INFN sezione di Roma I and Istituto Superiore di Sanit\`a, 00161 Roma, Italy\\
$^8$ Institute for Nuclear Research, 117312 Moscow, Russia\\
$^9$ INFN Laboratori Nazionali di Frascati, 00044 Frascati, Italy\\
$^{10}$ RRC ``Kurchatov Institute", 123182 Moscow, Russia}

\end{center}

\vspace{0.2in}

{\bf Abstract} - In this contribution, we will present our first preliminary 
measurement of the differential cross section for the reaction 
$\gamma n \rightarrow \eta n$. Comparison of the reactions 
$\gamma p \rightarrow \eta p$ for free and bound proton 
(D$_2$ target) will also be discussed.

\section{Introduction}

In our attempt to extract the properties of excited states of the nucleon, meson photoproduction on the neutron is of utmost importance in bringing complementary information, hence additional constraints on the theoretical interpretations. After having studied extensively meson photoproduction on a proton target, we have recently started looking at reactions on the neutron using data taken with a deuteron target. As for the proton~\cite{aja98,ren02}, one of our main goal was to study eta photoproduction by measuring both the differential cross section and the beam asymmetry $\Sigma$. This combination will allow not only to better fix the parameters of the dominant S$_{11}$(1535) but also to explore the nature of the other contributions and test the validity of the Moorhouse selection rule.

Moreover, this channel has recently drawn much attention in connexion with the pentaquark $\theta^+$. This state, first observed in 2003 by the LEPS collaboration~\cite{nak03}, still awaits a definite confirmation as discussed during the first session of this conference.
The chiral soliton model $\chi$SM~\cite{dya97} which is at the origin of this discovery, actually predicts the existence of an entire new anti-decuplet, the $\theta^+$ being its lightest element. Observation of these other states would provide decisive information for the validity of this model. Of particular interest to us is the second element of the anti-decuplet, identified as the P$_{11}$(1710) resonance in the original version of the model. This non-strange resonance should have a mass around 1700~MeV and a width $\sim$10~MeV, strikingly narrower than usual resonances. This state has been predicted by Polyakov and Rathke~\cite{pol03} to couple preferentially to the neutron. Besides eta as well as kaon photoproduction have been suggested as particularly sensitive channels, both of them being accessible to the GRAAL facility.

K$^0$ and K$^+$ photoproduction on the neutron are two reactions with a very low cross section combined to an intricate final state for our set-up. Even by summing up all available data, we have a limited statistics and cannot make any definitive statement for the time being. By contrast, eta photoproduction on the neutron is a rather "easy" channel for GRAAL for which the measurement of the differential cross section is a realistic objective. This information will allow to investigate the nature of new contributions, exotic or not, through Partial Wave Analysis.

\section{The GRAAL facility}

The GRAAL facility uses a tagged and polarized $\gamma$-ray beam produced by Compton scattering of laser light off the 6.03~GeV electrons circulating in the storage ring of the ESRF (Grenoble, France). The tagged energy spectrum ranges from 600 to 1500~MeV. Data dicussed hereafter have been obtained with highly polarized linear photons (P$_l \geq$50\% over the whole range).

The non-magnetic 4$\pi$ detector LAGRANGE has the nice property to detect all neutral and charged particles over almost the full angular acceptance. It is composed of three layers: MWPC's for the tracking of charged particles; thin plastic scintillators for charged particles identification; and a third layer for calorimetry with a BGO ball, made up of 480 crystals, for the detection of $\gamma$-rays in the central region ($\theta \geq 25^0$) and a shower wall to cover the forward region. This latter detector posseses a high efficiency and a good angular resolution for photons but provides no energy measurement. Both detectors can detect neutrons with a good efficiency (respectively $\sim$40\% and $\sim$20\%), the shower wall giving in addition n/$\gamma$ identification thanks to its ToF measurement.

\section{Analysis procedure}

When going from a free nucleon to a nucleon bound in a deuteron target, one has to take into account nuclear effects : the Fermi motion of the struck nucleon and possible final-state interactions. Hence, in order to extract any meaningful information on a reaction that occured on a bound neutron, one has to first evaluate these nuclear effects on the proton, by comparing the free and bound proton differential cross sections. At a later stage, one should be able with the collaboration of theoreticians to extract the free proton cross section from the bound one, and apply similar corrections on the bound neutron. 

In this work, our goal was to extract simultaneously the differential cross sections for the reactions $\gamma p \rightarrow \eta p$ and $\gamma n \rightarrow \eta n$, both from the deuteron target. The analysis procedure we have followed was identical to the one used for the free proton with, in addition, a veto on the recoiling spectator nucleon. Selection of the reaction was obtained by means of the $\eta$ invariant mass and using the two-body kinematics, assuming the struck nucleon at rest. For this first attempt and in order to limit uncertainties arising from neutron efficiency and identification, we have restricted the detection of neutrons at forward angles ($\theta \leq 25^0$). The shower wall response has been simulated for neutrons and its efficiency is estimated to be around 22~\%. The BGO efficiency is under study and could be as high as 40-50\%.

All distributions look very similar to the free case and, because of Fermi motion, are slightly broadened. Despite the fact we are measuring all kinematical variables (angles and energy) of outgoing particles (2 $\gamma$ and n or p), we cannot precisely measure the Fermi momentum of the struck nucleon, our resolution being of the same order of magnitude. Simulation studies have shown that the application of narrow cuts based on the two-body kinematics can eliminate the largest Fermi momenta and therefore slightly reduce the broadening effect ($\sim$- 30\%). Nevertheless, such cuts make it difficult to control the analysis efficiency and hence to get reliable cross section. 

As an illustration, the $\eta$ invariant mass is displayed in Fig.1 for both reactions.
As can be readily seen from the tails of these distributions, the level of continuous background is very low ($\leq$2\%). An evaluation from a Monte-Carlo simulation of contamination from neighbour reactions gives similar results ($\eta$N$\leq$1\%, $\eta \pi$N$\leq$1-2\%). Furthermore, cross sections have been checked to remain stable when changing the width of cuts, confirming the absence of any significant background.

\begin{figure}
\begin{center}
\includegraphics[width=8cm]{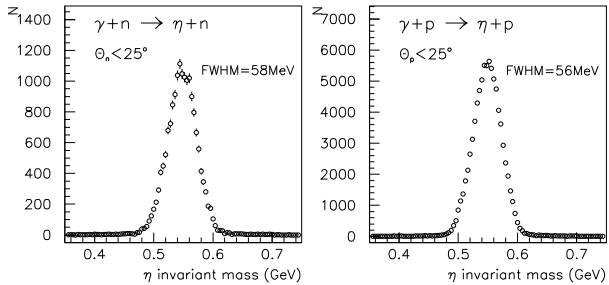}
\vspace*{8pt}
\caption{Eta invariant mass for the reaction $\gamma n \rightarrow \eta n$ (left) and
$\gamma p \rightarrow \eta p$ (right) from a deuteron target.}
\end{center}
\end{figure}

\section{Comparison between free and bound proton}

Comparison of differential cross sections for the reaction $\gamma p \rightarrow \eta p$ for the free and bound proton are displayed on Fig.~2 at four CM $\eta$ angles versus E$_{\gamma}$. These results are preliminary and error bars are only statistical. Because of the selection of neutron/proton at forward angle in the present analysis, the $\eta$ angular range is limited to angles larger than 90$^0$. The curves represent the SAID FA02 solution with (dashed line) and without (solid line) Fermi motion convolution. At backward angles (cos($\theta^{\eta}_{CM}$)=-0.95 and -0.75), the moderate difference between the free and the bound proton is fairly consistent with the Fermi motion broadening of the S$_{11}$(1535) peak. 
This is not the case for the most forward bin (cos($\theta^{\eta}_{CM}$)=-0.35, $\theta^{\eta}_{CM}$=110$^0$) where a slight discrepancy is seen between the free and the bound proton, incompatible with Fermi motion. In any case, the good overall agreement is again a good indication that no sizeable background is present. Extension of this measurement over the full angular range is under way and will help to understand whether this is the signature of some nuclear effect or some analysis bias. 

\begin{figure}
\begin{center}
\includegraphics[width=8cm]{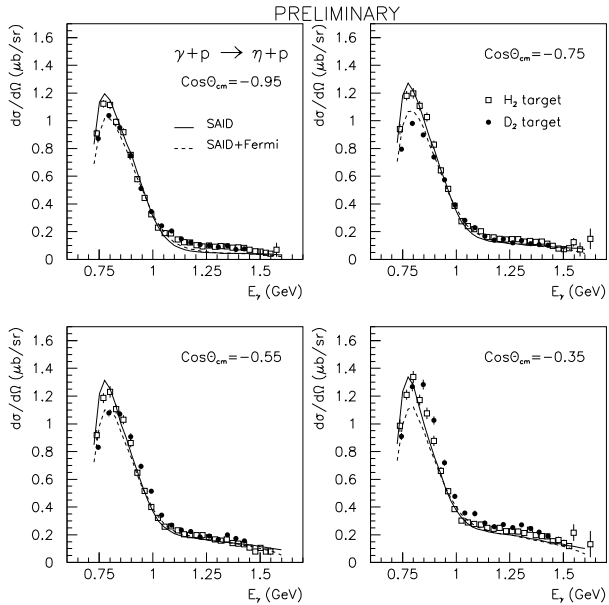}
\vspace*{8pt}
\caption{Differential cross section for the reaction $\gamma p \rightarrow \eta p$ at four $\eta$ CM angles versus E$_{\gamma}$. Comparison between the free (open squares) and bound (close circles) proton. The curves represent the SAID FA02 solution with (dashed line) and without (solid line) Fermi motion convolution.}
\end{center}
\end{figure}

\section{Differential cross section for the reaction $\gamma n \rightarrow \eta n$}

\begin{figure}
\begin{center}
\includegraphics[width=8cm]{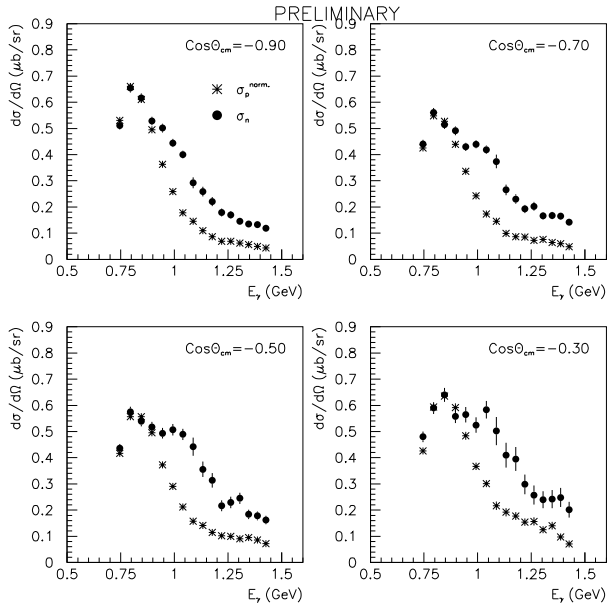}
\vspace*{8pt}
\caption{Differential cross section for the reaction $\gamma n \rightarrow \eta n$ at four CM $\eta$ angles versus E$_{\gamma}$ (close circles). For comparison, the renormalized cross section for $\gamma p \rightarrow \eta p$ on the bound proton (stars) is also plotted.}
\end{center}
\end{figure}

Fig.~3 displays the differential cross section of the $\gamma n \rightarrow \eta n$ reaction at four CM $\eta$ angles versus E$_{\gamma}$. It is compared to the $\gamma p \rightarrow \eta p$ for the {\em bound} proton, normalized to the neutron cross section in the S$_{11}$(1535) region. These two results have been obtained using identical procedures and it is likely that the bound neutron cross section suffers the same ``effect" as the bound proton close to 90$^0$. 

The shape of both cross sections is exactly identical below 900~MeV, indicating the same dominance of the S$_{11}$(1535) resonance on the neutron as for the proton close to threshold. 
The measured ratio $\sigma_n/\sigma_p$ is around 0.6, in fair agreement with previous results.\cite{hof97}

Above 900~MeV, whereas the proton falls off rapidly, a clear structure appears on the neutron with a marked angular dependence, evolving from a shoulder at backward angle (top-left) to a peak-like structure close to 90$^0$ (bottom-right) at E$_{\gamma}\approx$1~GeV, i.e. W$\approx$1.7~GeV.

An other interesting and complementary comparison is displayed on Fig.~4. The cross sections ($\eta$n and normalized $\eta$p) of Fig.~3, integrated over the angular range where the free and bound proton cross sections are consistent with Fermi motion widening, are displayed on the left-hand side, whereas the $\eta$N invariant mass is plotted on the right-hand side. This latter variable is calculated using only final state information and is therefore ``free" of Fermi motion. In other words, the broadening of a narrow structure would be only due to the resolution of our apparatus. The two distributions exhibit a similar behaviour with a resonant-like structure around W=1.7~GeV.

In the context of the search for the non-strange member of the $\chi$SM anti-decuplet discussed above, this seemingly resonant structure is of great interest. Yet, it is not a narrow one which would certainly be the signature of something ``exotic", be it a pentaquark or not. As mentioned previously, because of Fermi motion, our choice to extract cross section makes it impossible to observe such a narrow state in the present analysis. The simulation tells us that a 10~MeV broad resonance would become $\sim$130~MeV wide in E$_{\gamma}$ or $\sim$70~MeV in ($\eta$n) invariant mass. Even by applying stringent cuts, one may only moderatly reduce the width and we want therefore to further explore the possibility to minimize the effect of Fermi motion by using alternative analysis methods.

On the other hand, the observed structure could be compatible with a usual broad resonance, several candidates being in this energy range: S$_{11}$(1650), D$_{13}$(1700) or D$_{15}$(1675). Like for the proton case, the beam asymmetry $\Sigma$ will bring valuable information to better constrain PWA and to discriminate among these various possibilities. We have now obtained preliminary values for $\Sigma$ over the full angular range.\cite{rac04} Most surprisingly, the overall shape is rather similar between the proton and the neutron, in sharp contrast with the cross sections.
 
\section{Conclusions}

\begin{figure}
\begin{center}
\includegraphics[width=8cm]{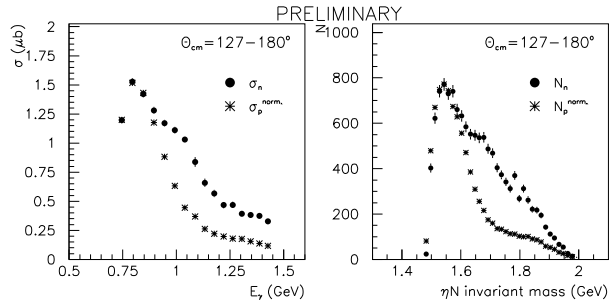}
\vspace*{8pt}
\caption{Left: Differential cross section integrated over the covered angular range (113-180$^0$); neutron (close circles) compared to renormalized proton (stars). Right: Idem as left for invariant mass of ($\eta$n, p) calculated from final state information.}
\end{center}
\end{figure}

In summary, we have presented preliminary results for the differential cross sections of $\gamma n \rightarrow \eta n$ using data taken with a deuteron target. Results for the reaction $\gamma p \rightarrow \eta p$ have been also extracted from the same set of data. We confirm that the reaction on the neutron is dominated by the S$_{11}$(1535) close to threshold. By contrast, a resonant-like structure is observed on the neutron around W=1.7~GeV, not seen on the proton.

\end{normalsize}

\end{document}